\documentclass[twocolumn,12pt]{article}
\usepackage{amsmath}            
\usepackage{amssymb}            
\usepackage{wrapfig}            
\usepackage[latin1]{inputenc}   
\usepackage{fancyhdr}           
\usepackage{lscape}             
\usepackage{hyperref}           
\usepackage{capt-of}            
\usepackage{makeidx}            
\usepackage{ifpdf}

\ifpdf
\usepackage[pdftex,final]{graphicx}
\else
\usepackage[dvips]{graphicx}
\fi

\begin{document}
\pagestyle{plain}

\twocolumn[{\csname @twocolumnfalse\endcsname

\begin{center}
\underline{Roberto Guida, Alessandro Cacciani} \\
\textbf{Observation of the atmospheric sodium layer with a magneto-optical filter (MOF)} \\
Department of physics, University of Rome ``La Sapienza'' \\
Rome (Italy) \\
roberto.guida@icra.it \\
\end{center}

\begin{abstract}
The Mesosphere is interested by important chemical and dynamical phenomena, so observation of its Sodium layer's behavior has became a common target of several research plans all over the world. In order to study its dynamical and chemical variation during daytime (so the solar flux, which is not localized, stimulates the Sodium emission) we want to observe from the space continuously the Sodium emission (and perform differential analysis looking for periodic changes), using a small telescope with a Magneto-Optical Filter and a image sensor. 
\end{abstract}

}]


\section{\underline{THE OBSERVABLE}}
\begin{figure}[h]
\centering
\includegraphics[width=8 cm]{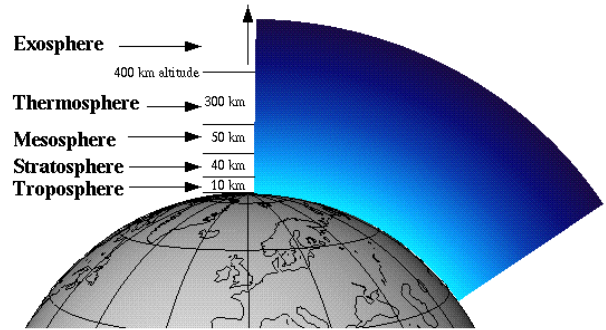}
\caption{Relative altitudes of various components}
\label{Fig1}
\end{figure}

\begin{figure}[h]
\centering
\includegraphics[width=8 cm]{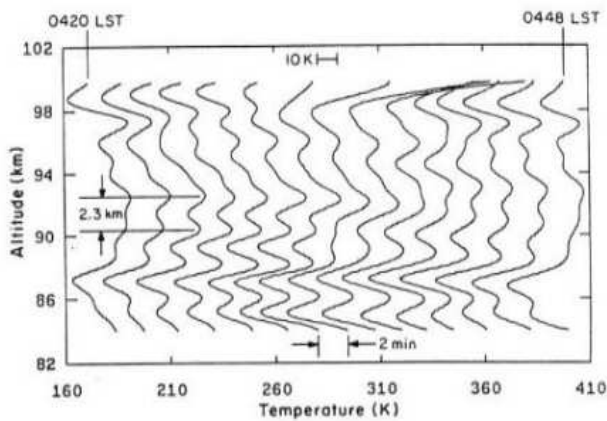}
\caption{Temporal series of temperature profiles}
\label{Fig2}
\end{figure}

\begin{figure}[h]
\centering
\includegraphics[width=8 cm]{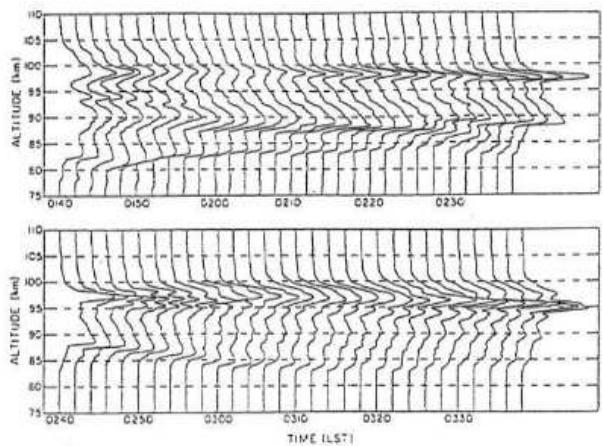}
\caption{Density profiles}
\label{Fig3}
\end{figure}

The atmosphere of the Earth contains a layer of metallic Sodium. Even though almost 30 years have passed since the first routine investigations by Gibson, Standford and Bowman have made \cite{Bow}, we still have no definite answers about its origin, large scale properties and temporal evolution. This lack of knowledge is the basic reason why several research groups, all around the world, are working on this field and, definitely, constitutes the main trigger for our proposal.\\Neutral atoms of Sodium have been first discovered by Slipher in 1929 \cite{Sli} and they are located between $80$ and $105$~Km in the Mesosphere (Fig. \ref{Fig1}).
The peak density of this Sodium layer is around $90$ Km at a level of $10^9\div10^{10}$ atoms m$^{-3}$ so that its investigation faces some difficulties as follows:
\begin{itemize}
\item Its altitude is not suitable for balloons (too high) as well as for orbiting satellites (too low).
\item The background solar daylight radiation, diffused by the atmosphere is far exceeding its faint yellow glow, leading to the problem of the signal to albedo noise, discussed in section 3.1.
\end{itemize}
As far as the first point, \textsc{lidar}s (systems composed by a powerful laser which excites the atmosphere, coupled with a fairly large telescope to collect the backscattered radiation from the various component) have made possible accurate observations, limited however to point-like targets only, so precluding large scale field of view (FOV) and spatio-temporal studies of its evolution.\\As far as the second point, routine observations are usually being performed at night time only, so precluding long and uninterrupted temporal series to get better frequency spectra of its variations. Attempt have been made during the day, reducing the divergence of the \textsc{lidar} beam or using narrow band Fabry-Perot filters that, however, are affected by extreme thermal and mechanical instability.\\In spite of the above difficulties, the Mesospheric Sodium is the most frequently studied metallic component of the Earth's atmosphere because of its high backscattering cross section, $7\cdot10^{-17}$~m$^2$ ster$^{-1}$, and abundance, $3\cdot10^{13}$ atoms per m$^2$ \cite{Hum}.\\However, so far, we are not able to respond with certainty to the basic question of the its origin. In 1939 Chapman \cite{Chap} described the characteristic and listed some possible groups of Sodium sources:
\begin{itemize}
\item Terrestrial sources like volcanic dusts projected up to $10\div30$ Km and salt particles brought from the oceans by updrafts.
\item Extraterrestrial sources like meteorites and comet dusts, gas from the sun, and interstellar Sodium that the Earth collect during its movement in the Solar System.
\end{itemize}
Most likely, ablation of meteorites and comet dusts is the main source mechanism; however, difficulties arise when trying to justify the observed abundance.\\Moreover, researchers find that the Sodium layer is changing as a function of geographical coordinates, seasons and even time during the day. Complex chemical reactions are taking place that involve many ions and neutral particles present at high levels: all depending from the basic parameter governing the evolution, that is the temperature (Fig. \ref{Fig2}).
Finally, one important aspect is the dynamical behavior, driven by the tidal effects and gravity waves.\\These effects, with some contribution of atmospheric turbulence, are mainly responsible for short periods evolution. The available data \cite{Carla} show oscillations of the Sodium layer: in particular, the density and the thickness oscillate in phase, while the average altitude is 180 degrees out of phase. Moreover, sometimes, sporadic layers have been detected with very high Sodium density and short formation time, whose origin is still not understood (Fig. \ref{Fig3}).
\\Because of its high resonance cross section, the Mesospheric Sodium constitutes a very good mark able to reveal waves and short duration transients phenomena in the Earth atmosphere. For this reason it could be very important to image and to record such events in order to:
\begin{itemize}
\item Further clarify the source of Sodium from meteorites and comet dusts detecting its consequent density variation. Since meteoric events are randomly localized in space and time, we would need a continuous monitoring of a large FOV, obtainable with measures from the space.
\item Get evidence and possibly obtain spatio-temporal Fourier analysis of both tidal and gravity waves.
\item Try to combine our space observations with other techniques from the ground, for example using data available from \textsc{lidar} routine measurements.
\end{itemize}
\section{\underline{INSTRUMENT}\\\underline{DESCRIPTION}}
\begin{figure}[h]
\centering
\includegraphics[width=8 cm]{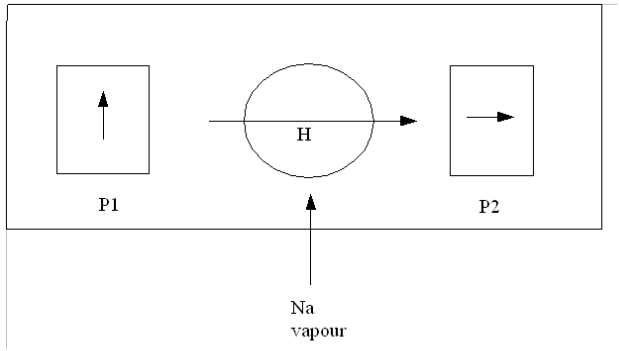}
\caption{Schematic diagram of the MOF}
\label{Fig4}
\end{figure}

\begin{figure}[h]
\centering
\includegraphics[width=7.5 cm]{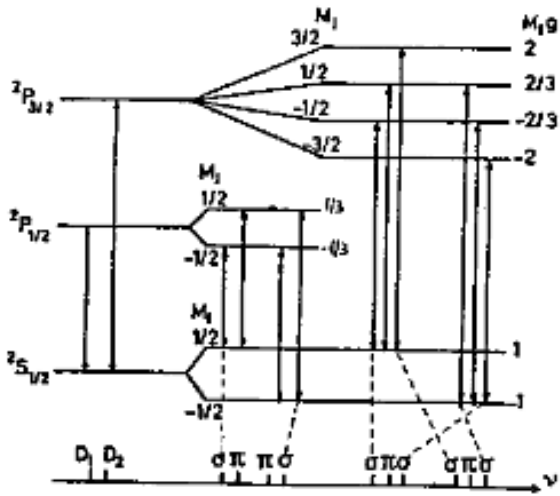}
\caption{Zeeman effect}
\label{Fig5}
\end{figure}

\begin{figure}[h]
\centering
\includegraphics[width=8 cm]{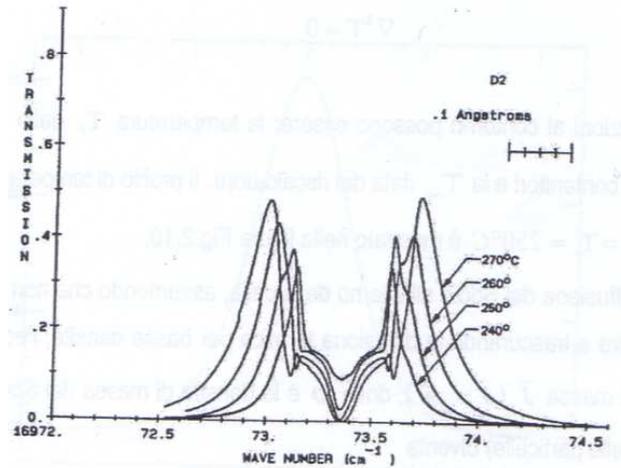}
\caption{Transmission profiles at different temperatures, with H=2000 Gauss.}
\label{Fig6}
\end{figure}

The Magneto-Optical Filter (MOF) is a particular instrument that gives a really narrow bandwidth, an high transmission (almost 50\%) and a good stability. It can work only in a small range of wavelengths, well defined, like Na and K doublet, and can be thought as an high resolution spectrograph within those wavelengths.\\It has been developed by professor Alessandro Cacciani of the Rome university ``La Sapienza'' at the end of '60 years and its main use has been for Sun's study; during the last years it has been used for the analysis of Jupiter's oscillation modes and for the observation of mesosphere with \textsc{lidar} technique.\\It is made of two separated unit: a MOF and a Wing Selector (WS).\\The MOF is made by two crossed polarizers (P$_1$, P$_2$) and a metallic vapour, in this case sodium, between them in a longitudinal magnetic field ($2000\div4000$ Gauss). The scheme of the filter is shown in Fig. \ref{Fig4}.
The working principle of the MOF is based on two concurrent effects, namely the Zeeman effect in absorption as shown in Fig. \ref{Fig5} and a sort of Faraday rotation close to the line's wings called Macaluso-Corbino effect. Both change the polarization in and around the resonance lines, leading to the typical total transmission profile shown in Fig. \ref{Fig6}.
\\In fact the light coming from the left crosses the first polarizer and becomes so linearly polarized; consider then the sodium vapor in a magnetic field of few K Gauss: following the Zeeman rules, the vapor absorbs and ri-emits two circularly polarized wavelengths ($\lambda_{D1}=5895.92$ $\AA$ and $\lambda_{D2}=5889.95$ $\AA$). All the other wavelengths are still linear polarizated due to the first polarizer and so they will be stopped by the last polarizer orthogonal to the first one, while the two close wavelengths can pass due to their circular polarization: so we get a couple narrow bands that can be close at will depending on the magnetic field strength; in fact lowering the magnetic field lower the Zeeman splitting of the atomic levels.\\The wing selector (WS) is also made of a cell containing sodium vapor in a more strong magnetic field; before this cell there is a quarter-wave plate. The WS deletes alternatively one of the two bands transmit from the MOF, depending by the orientation of the quarter-wave plate. The wing selector is particularly used in Solar physics, to measure the velocities of the different parts of the Sun.
\section{\underline{THE TECNIQUE}\\\underline{FROM SPACE}}
\begin{figure}[h]
\centering
\includegraphics[width=7 cm]{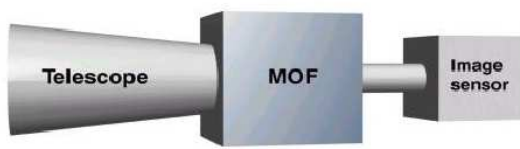}
\caption{Mechanical assembly of the instrumentation}
\label{Fig7}
\end{figure}

Referring to Fig. \ref{Fig7} where our instrument is shown as composed by three parts (a telescope, the MOF filter and an image sensor), we want to stress here the importance and the characteristics of the MOF filter for space use.
Its working principle has been described previously, and summarizing, it is a very stable and narrow band filter, about $50\cdot10^{-3}$ $\AA$, and it achieves unsurpassed performances of perfect tuning in the core of the Sodium yellow lines. Its weight is about 1 Kg and the dimensions are $10\times10\times10$ cm$^3$ making it so very attractive for space applications. As a filter, it will be located between the telescope and the image sensor so that this last can only see light within about 50 milliAngstrom centered in the Sodium emission wavelength values. In this manner we are able to reject all the other wavelengths of the solar spectrum so that we can definitely say that the MOF produces an artificial night which is the necessary condition to detect the faint yellow glow originated by the mesospheric Sodium. The most important consequence of this artificial night is that we can work on daylight, using the solar radiation, instead of the uncomfortable \textsc{lidar} apparatus, to get the Sodium excited; moreover, the solar radiation is not localized, as the laser beam is, and so we can observe many points at once of a large bi dimensional FOV. A drawback of this otherwise ideal situation is that we must compete with the albedo radiation from the Earth inside the same wavelength pass-band of the MOF. This problem is discussed in the following section devoted to the signal-to-noise ratio.
\subsection{\underline{Signal-to-Noise ratio}}
\begin{figure}[h]
\centering
\includegraphics[width=8.3 cm]{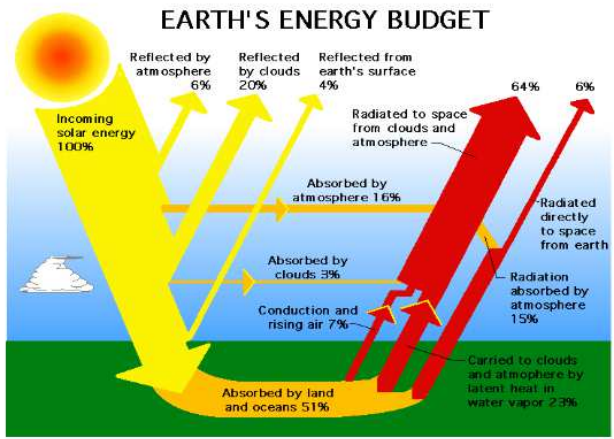}
\caption{Earth's energy budget}
\label{Fig8}
\end{figure}

Here we need to compute the expected signal and the competing luminosity from the Earth: as far as the expected signal, the latter comes from the Sodium atoms. We have:
\begin{itemize}
\item The average value of the Sodium columnar abundance that is around $3\cdot10^{13}$ atoms m$^2$ \cite{Hum}.  
\item The cross section for resonance scattering that is around $7\cdot10^{-17}$ m$^2$ ster$^{-1}$ \cite{Hum}.
\end{itemize}
The product of them gives the probability that one single solar photon is scattered back from the Sodium layer inside one steradiant. This amounts to $2\cdot10^{-3}$ events per second and per steradiant.\\We need to compare this number with a similar probability coming from the other components of the Earth, namely: the albedo from the clouds, the oceans, the lands and the atmosphere. Each component displays a different albedo luminosity (clouds are white and oceans are blue), therefore we want to select the observations in the most favorable condition for the yellow part of the spectrum, that is, above the oceans and the air as a background. Unfortunately, albedo data in the narrow MOF band width are not available: only integral or wide band spectral resolution data have been estimated in the whole solid angle (Fig. \ref{Fig8}).
It is, therefore, possible that most of the measured yellow albedo from the oceans or the air is coming from the high resonance radiation of the Sodium atoms (which is our signal).\\To give some numbers, let us take the yellow albedo of the air (with the exclusion of the Sodium resonance) to be of the order of $10^{-2}$, that is less than $10^{-3}$ in a single steradiant. This would mean that the above quoted resonance probability is larger than the albedo in the same spectral bandwidth. We add that at the terminator (the twilight zone) the Sodium layer at $90$ Km is excited by the solar radiation while the Earth's atmosphere is not completely illuminated: so that the situation is even more favorable.
\subsection{\underline{Data analysis}}
The first order data analysis is to perform suitable time lag differences in order to get evidence of transient events, specially during the times of Leonids, Perseids, etc. In this manner we can visualize the location and geometrical configuration of any glow increase. The temporal coincidence with the periods of known meteor swarms, will witness about their credibility. Then we can try to evaluate the consequent density variation in the nearby Sodium regions. Depending on the system sensitivity to small variations we can also try to get evidence of waves and look at spatio-temporal power spectra of the observed surface oscillations as in solar seismology where in this way people can make good diagnostics of the deeper layers of the star.\\This kind of information, if prolonged in time, we think are surely useful also for other disciplines connected with the very important problem of global changes.

\end{document}